\documentstyle[epsfig,astrobib]{mn-ab}

\title[Near infrared galaxy survey]
  {Galaxy number counts- IV. surveying the Herschel deep field in the near-infrared}

\author[McCracken et.\ al]
{H.~J.~McCracken$^{1,5}$, N. Metcalfe$^1$, T. Shanks$^1$, A. Campos$^{1,2}$,
\newauthor J. P. Gardner$^{1,3,4}$, R. Fong$^1$\\ 
$^1$Department of Physics, University of Durham Science Laboratories, South Rd, Durham DH1 3LE.\\
$^2$ Instituto de Matematicas y Fisica Fundamental (CSIC)
Serrano 113 bis,  E-28006 Madrid, Spain \\
$^3$ Laboratory for Astronomy and Solar Physics, Code 681, Goddard Space Flight Centre, Greenbelt MD 20771, USA\\
$^4$ NOAO Research Associate.\\
$^5$ Present address: Laboratoire d'Astronomie Spatiale, 13376 Marseille Cedex 12, France\\}

\begin{document}
\def\sizex{12.0 cm}
\def\bigx{16.0 cm}
\def\sizesmallx{8.0 cm}
\maketitle

\begin{abstract}
  We present results from two new near-infrared imaging surveys. One
  survey covers $47.2$ arcmin$^2$ to $K(3\sigma)=20^m$ whilst a second,
  deeper catalogue covers a sub-area of $1.8$ arcmin$^2$ to
  $K(3\sigma)=22.75^m$. Over the entire area we have extremely deep
  optical photometry in four bandpasses ($UBRI$), allowing us to track
  the colour evolution of galaxies to very faint magnitude limits.
  
  Our $K-$ band number counts are consistent with the predictions of
  non-evolving models with $0\leq q_0 \leq 0.5$. The $K-$ selected
  $(B-K)$ galaxy colour distributions from our surveys move sharply
  bluewards fainter than $K\sim20$. At brighter magnitudes ($K<20^m$)
  our $K-$ selected $(B-K)$ distributions indicate a deficiency of red,
  early-type galaxies at $z\sim1$ compared to the predictions of
  passively evolving models, which implies either a pure luminosity
  evolution (PLE) model where star-formation continues at a low level
  after an initial burst, or dynamical merging. At fainter magnitudes,
  the continuing bluewards trend observed in $(B-K)$ can be explained
  purely in terms of passively evolving PLE models.  We detect $0.5\pm
  0.1$ galaxies arcmin$^{-2}$ with $(I-K)>4$ and $19<K<20$~mag.
  Although this is a factor of $\sim3$ ($2\sigma$) more objects than
  the recent survey of \citeANP{AB} this is still lower than the
  predictions of standard passively evolving models and more consistent
  with PLE models containing small amounts of on-going star-formation.
  Our observed numbers of $(I-K)>4$ galaxies at $K\sim20$ are lower
  than the predictions of passively evolving models or PLE models with
  a low level of continuing star-formation, suggesting that at least
  part of the larger deficiency observed in $(B-K)$ at $K\sim20$ may be
  due to star-formation rather than dynamical merging.
  
  As we and others have noted, the number redshift distribution at
  $18<K<19$ of recent, deep $K-$ selected redshift surveys is well
  fitted by non-evolving models, and passively evolving models with a
  Salpeter or Scalo initial mass functions predict too many galaxies
  with $z>1$.  Dynamical merging is one possible solution to reduce the
  numbers of these galaxies but (as we have suggested previously) a
  dwarf-dominated IMF for early-type galaxies could offer an
  alternative explanation; we show here that such a model reproduces
  well the optical-infrared colour distributions and $K-$ band galaxy
  counts.

\end{abstract}

  

\begin{keywords}
 Galaxies\ : evolution\ --Galaxies:photometry.
\end{keywords}

\section{Introduction}
\label{sec:Introduction}

In our previous papers \cite{1991MNRAS.249..498M,MSFR,MSCFG} we have
investigated the evolution of the counts and colours of faint field
galaxies, culminating in extremely deep optical imaging of a
representative region of sky we call the ``Herschel Deep Field''
\cite{MSF}.  In this paper we present new near-infrared observations of
this field, and combine this data with our existing ultra-deep optical
imagery.  Near-infrared observations have many advantages which are
oft-quoted in the literature, among them, the insensitivity of the $k-$
correction to type or star-formation history at intermediate redshifts
is perhaps the most relevant in a cosmological setting.

Previously we have considered our observations in the context of pure
luminosity evolution (PLE) models.  Recently, additional support for
these models has come from fully complete, blue-selected, redshift
surveys \cite{CSH2} which have revealed the presence of an extended
tail in the redshift distribution, partially resolving the paradox that
non-evolving galaxy formation models could be used to fit the results
of the early faint redshift surveys \cite{1995MNRAS.275..169G,BEG}. Such
models, as we have shown, are able to reproduce all the observable
quantities of the faint field galaxy population (counts, colours,
redshift distributions) at least for low $\Omega_0$ Universes and
within current observational uncertainties.  High $\Omega_0$ Universes
can be accommodated by the model if we add an extra population of low
luminosity galaxies with constant star-formation rates which boost the
counts at faint ($B>25^m$) magnitude levels \cite{MSCFG,1997ApJ...488..606C}.
In the PLE model, galaxies form monolithically at high redshift.
Changes in luminosity and colour after the formation event are modelled
with stellar population synthesis models. $K-$ selected galaxy samples,
at least until $K~\sim~20$, are dominated by early-type galaxies;
therefore, changes in optical-infrared galaxy colours are particularly
sensitive to evolution of this galaxy class. By investigating
near-infrared number counts and colour distributions we can potentially
address questions concerning the formation and evolution of elliptical
galaxies.

\begin{table}
\begin{center}
\caption{Photometric Limits of the Herschel Deep Field.}
\begin{tabular}{lccccc}
Filter& U & B & R & I & K \\
Limit (3$\sigma$) & 26.8 & 27.9/28.2 &26.3& 25.6&20.0/22.75\\
Area (arcmin$^2$) &  46.4 & 46.8/2.8 & 48.5 &52.2  & 47.2/1.78\\
\end{tabular}
\end{center}
\end{table}
\label{tab:intro.limits}

The two surveys presented here are complementary.  The ``wide'' survey
covers a large, $\sim50$ arcmin$^2$ area to a shallow,
$K(3\sigma)=20^m$, depth, whilst the ``deep'' survey covers a much
smaller area of $\sim 1.8$ arcmin$^2$ to a much fainter
$K(3\sigma)=22.75^m$ limiting magnitude. In combination, both surveys
provide a $K-$ limited sample extending from $K\sim16^m$ to $K=22^m.75$.

This paper is organised as follows: Section~\ref{sec:ObsDat} describes
the data reduction techniques used in both the wide and the deep
surveys; in Section~\ref{sec:Results} number counts and colour
distributions for our both data sets are presented; in
Section~\ref{sec:Interpr-Modell} we discuss these results in terms of
our modelling procedure; and finally, in Section~\ref{sec:IntDisc} we
summarise the main conclusions we draw from our work.

\section{Observations and Data Reductions}
\label{sec:ObsDat}

\subsection{Reducing the ``Wide'' Survey}
\label{sec:RedWid}

The observations were made over four nights in July 1995 at the 3.8m
United Kingdom Infrared Telescope (UKIRT) using the IRCAM3
near-infrared detector with a standard $K-$ ($2.2~\mu$m) filter.
Conditions were photometric on two of the four nights. The mean seeing
was $\sim 1''$ FWHM\@.  The centre of the region we imaged, RA $0$h
$19$m $59$s Dec $+0^o$ $4'$ $0''$, corresponds to the centre of our
deep optical fields which are described in \citeN{MSF}. For this region
we have imaging data complete to the limiting magnitudes listed in
Table~\ref{tab:intro.limits}. The optical observations were made at the
4.2m William Herschel and the 2.5m Isaac Newton telescopes (WHT and INT
hereafter) and cover approximately $\sim 50$ arcmin$^2$. 

The IRCAM3 detector, a $256 \times 256$ element InSb chip, has a pixel
scale of 0.286$''$ and consequently each image covers only $\sim 1.5$ \ 
arcmin$^2$.  For this reason we observed a mosaic of $6\times 6$ frames
which completely overlapped the previously-observed optical fields. The
entire mosaic was observed in each pass for two minutes at each
position, and offset slightly for each observation from the previous
pointing.  In total, the entire survey area was covered fifteen times,
giving a total integration time at each sub-raster of 30 minutes. The
short exposures for each image is a consequence of the high sky
brightness in the near-infrared and is fact the longest exposure time
possible with IRCAM3 in the $K-$ before sky background causes the
detector to saturate. Our data reduction technique for both infrared
surveys follows the methods outlined in \citeN{1995ApJS...98..441G}.

A two-pass procedure is adopted: for the first stage in the data
reductions a dark frame is subtracted from each night's images,
calculated from the median of the dark frames taken that particular
night. Next, the images are sorted in order of time of exposure. Then,
a {\it running sky\/} is constructed from the median of the six images
nearest in time to the frame being processed.  This is necessary as the
sky background varies on exceedingly short time-scales (i.e., under 30
minutes).  This running sky is then subtracted from the current image
and the procedure repeated for all the images in the mosaic. Next, for
each night a ``superflat'' is constructed by medianing together all the
observations for that night.  Each image is then divided by the
superflat.

Once all the images have been dark- and sky-subtracted and flat-fielded
they are grouped into 36 separate pointings and each stack averaged
together using the IRAF \textsc{imcombine} task. Next, masks were
constructed from each of the stacked images by flagging all pixels more
than $3\sigma$ from the sky background. These masks were then used in
re-calculating the running sky frames. This was necessary as bright
sources present in each of the individual exposures caused raised
counts in the original running sky frames. The whole procedure outlined
above was then repeated, using these new frames. 

Finally, the $36$ stacked sub-rasters were combined into a single image
by matching each sub-raster with objects on the extremely deep $B-$
band frame and using this to calculate the correct position of each
sub-raster relative to the others. The final image, after trimming,
covers an area of $47.2$ arcmin$^2$, overlapping our WHT deep optical
fields.

We calibrated this data by repeatedly ($\sim 30$ times) observing the
faint UKIRT Standards \cite{CH92} each night at a range of airmasses,
although fortuitously the mean airmass of the standards and of the
program field images was approximately the same, so accurate
determination of the extinction coefficient was not necessary. The rms
scatter between our standards measurements was $\pm 0.03$ magnitudes. 

\subsection{Reducing the ``Deep'' Survey}
\label{sec:RedDep}

Observations for the ``Deep'' survey were made over six nights in 1994
September at UKIRT using IRCAM3 with a $K-$ band filter. Conditions
were photometric during the first two nights, and the remainder of the
data were scaled to the photometry of bright objects in the field made
on those nights. The seeing ranged from 0.8$\arcsec$ to 1.4$\arcsec$.
The total observing time on the field was just over 100 ksec (27.9
hours), and was made up of 838 individual exposures of 120 seconds
each. Each individual exposure was made up of 12 co-adds of 10 seconds
each. The telescope was dithered using the offset guide camera in a
diamond pattern of 13 positions, each position separated by
5.25$\arcsec$ in both Right Ascension and declination. Data reduction
proceeded generally in the manner described in Section
\ref{sec:RedWid}. Dark frames were made several times each night, and
were subtracted from the images. A running sky was constructed, using
the median of 12 images (six before and six after). The running sky was
subtracted (after applying the two-pass object masking system described
above), and a ``superflat'' was constructed for each night and divided
into the data. The input dither pattern was recovered and adjusted by a
centroid of the brightest object in the field. The data taken each
night were shifted and co-added. The photometry of each night's data
were scaled to that of the photometric first two nights (which were
themselves consistent). The final image covers an area of 1.78
arcmin$^2$, and has a limiting depth of 3$\sigma$ = 22.75. The deep
field is centred at 00 19 59.6 +00 02 56 (1950.0) and covers 81$''$
square (140x140) 0.58'' pixels.

\begin{table*}
\begin{center}
\caption{Detection and measurement parameters used in both surveys.}
\begin{tabular}{lccccc}
  Frame&Limiting isophote&Detection limit&Min. r&Kron multiplying &Correction to\\
  &(mag/arcsec$^2$)&(K mag)&($''$)& factor&total (mag)\\
Deep K&24.50&23.50&1.25&1.40&0.32\\
Wide K&21.50&22.00&1.35&1.50&0.25\\
\end{tabular}
\end{center}
\label{tab:K.info}
\end{table*}

\subsection{Object Detection And Photometry}
\label{sec:ObjRed}

For both the deep and wide images we use the same prescription for
object detection and photometry as in our optical studies
\cite{MSFR,MSF}.  Note that to improve the reliability of image
detection both images were binned $2\times2$ (the $0.286''$ pixel size
of IRCAM3, combined with median seeing of $\sim 1''$ FWHM, means that
our raw images are oversampled).  We fit the sky background
approximately with a 3rd order polynomial and then subtract this from
the frame.  Deep isophotal image detection is then undertaken (to a
magnitude about $1^m$ deeper than the $3\sigma$ limit for the frame --
see Table ~\ref{tab:K.info}) and the objects removed from the frame and
replaced by a local sky value. \footnote{Our $3\sigma$ limit is
  defined as when the variance of the sum of the pixel values within our
  detection aperture reaches this value.} The resulting
image is heavily smoothed (with several passes with box smoothing up to
$10$ pixels on a side) and then subtracted from the original. The
result is an extremely flat image on which the isophotal detection is
then rerun.  At this stage, in order to reduce false detections, any
images within a few pixels of one another are recombined into one. A
Kron-type magnitude \cite{1980ApJS...43..305K} is then calculated for
each image using a local sky value.  This is essentially an aperture
magnitude calculated to $mR_k$, where $R_k$ is the Kron radius and $m$
is the multiplying factor given in Table~\ref{tab:K.info}. It is
necessary to set a minimum aperture equivalent to that for an
unresolved object, and apply a correction to total magnitude. These are
also listed in Table~\ref{tab:K.info}. A more detailed discussion of
this technique is given in \citeN{1991MNRAS.249..498M}.  Our
optical-infrared colours are measured inside fixed $1.5''$ radius
apertures. Note that to form the ``Deep'' $(B-K)$ colours, the $K-$
band magnitudes are matched with the $2'$ diameter, stacked, ultra-deep
$B-$ band image described in \citeN{MSF} (and see
Table~\ref{tab:intro.limits}).  All other colours are formed from
matches with the $7'\times7'$ Herschel Deep Field optical images.
Star-galaxy separation is taken from \citeN{MSF}.

\subsection{Confusion corrections}
\label{sec:Confus}
The deeper the ground-based number-counts are extended the more
important the corrections for confusion becomes. This is of particular
relevance for the deep K-band data. To make an estimate of the
completeness of our counts on this frame we have added numerous
artificial stars of various known magnitudes to the real data frame and
then reanalysed the resulting images using the normal data reduction
procedure. We then compare the measured magnitudes with those which we
input.  Table~\ref{tab:K.sim} gives the mean magnitudes, scatter, and
detection rate for these stars.  Note that an image is considered
undetected if it is merged with another image and the combined
brightness is a factor two or more greater than the true magnitude, or
if it is not found within $\pm2$ pixels of its true position.  As
expected the detection rates in the real data drop as the magnitude
falls.  This is almost entirely due to objects being merged with other,
brighter objects - isolated objects are recovered with almost $100\%$
efficiency. Note, however, that the mean recovered magnitudes are close
to the true values.  Although most objects at the limit of our data are
galaxies, they are in general close to being unresolved, and so we
correct our counts by the detection rates implied by our artificial
stars.

Due to the much wider mean separation between objects, the shallower
$K-$ band data do not suffer from confusion, and no correction is
made. 

\begin{table*}
\begin{center}
\caption{Results of adding artificial stars to the deep K data.}
\begin{tabular}{ccc}
True magnitude&Measured magnitude&Detection rate (\%)\\
\\
20.75&$20.70\pm0.17$&90\\
21.25&$21.22\pm0.21$&88\\
21.75&$21.76\pm0.26$&85\\
22.25&$22.36\pm0.36$&76\\
22.75&$22.83\pm0.37$&51\\
\end{tabular}
\end{center}
\label{tab:K.sim}
 \end{table*}

\begin{figure*}
\centering
\centerline{
\epsfysize = 18.5cm
\epsfbox{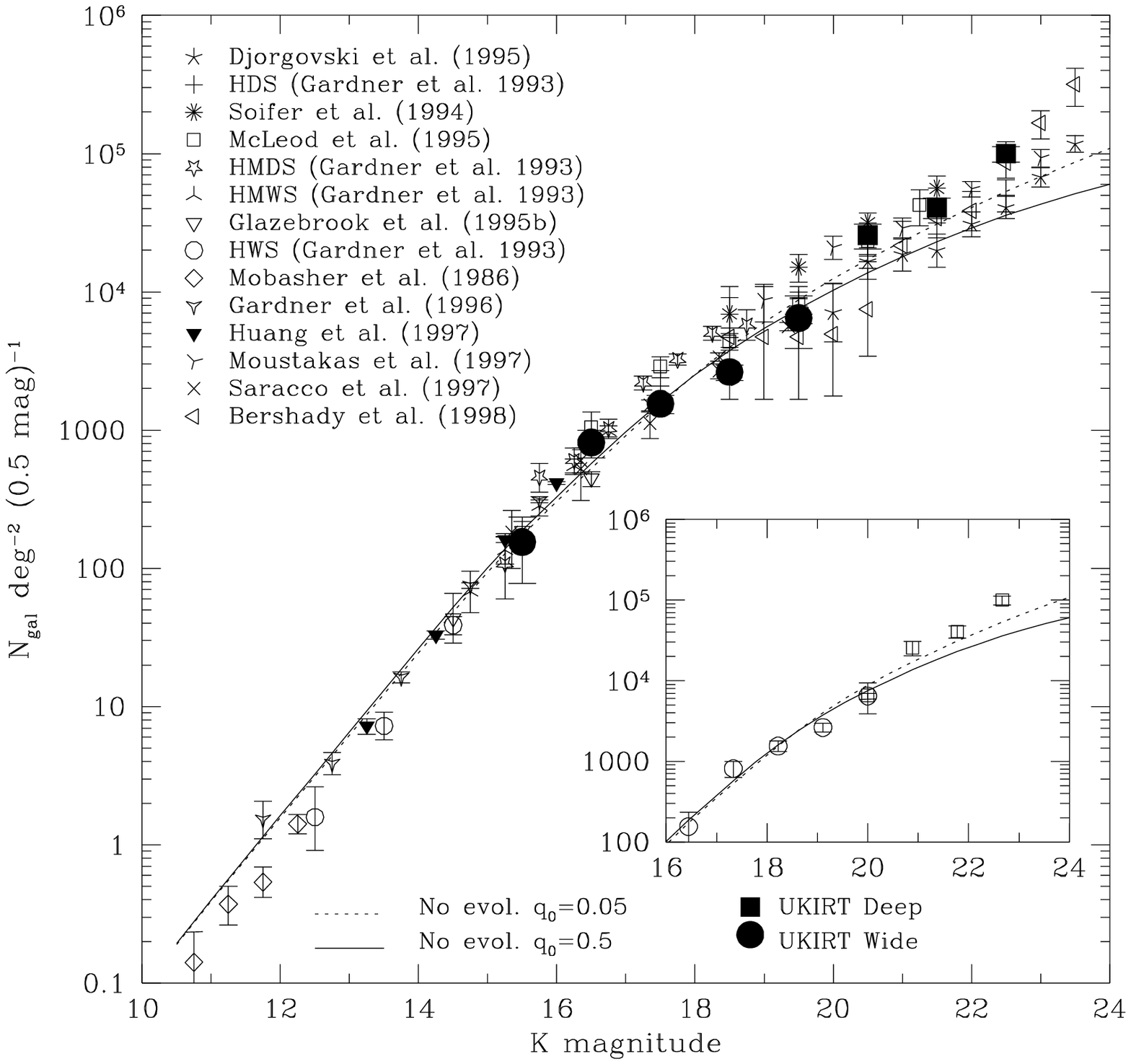}}
\vspace{-1.0in}
\caption{Differential galaxy number counts in half-magnitude intervals as a
  function of $K-$ limiting magnitude as measured from our UKIRT wide
  survey (filled circles) and from our UKIRT deep survey (filled
  squares). Also plotted is a compilation of published $K-$ band number
  counts and the predictions of the two non-evolving models described
  in the text for $q_0 = 0.05$ (dotted line) and $q_0=0.5$ (solid
  line).  Error bars are calculated from the number of galaxies in each
  magnitude bin using Poisson $\sqrt n$ counting statistics.  The inset
  shows the number counts from our surveys alone (this time with open
  symbols) compared to the non-evolving predictions.}
\label{fig:countscols.counts}
\end{figure*}
\nocite{DSP}
\nocite{1986MNRAS.223...11M}
\nocite{1995ApJS...96..117M}
\nocite{GSCF}
\nocite{1997ApJ...476...12H}
\section{Results}
\label{sec:Results}
The final ``wide'' catalogue contains 298 galaxies to a
$K(3\sigma)=20.0^m$ over an area of 47.2~arcmin$^2$. The deep catalogue
contains 86 galaxies to a $K(3\sigma) = 22.75^m$ over an area of
1.8~arcmin$^2$.

\subsection{$K-$ band Galaxy Number Counts}
\label{sec:countscols.Kband}

Number counts derived from both our catalogues are presented in
Figure~\ref{fig:countscols.counts}, which shows our data (with error
bars calculated using Poisson counting statistics) and a compilation of
published number counts. At bright magnitudes ($16^m < K^m < 20^m$) we
agree well with the counts from the literature, whereas at fainter
magnitudes ($20^m < K< 22^m$) our data favours the higher counts found
by \citeN{1998ApJ...505...50B} as opposed to the lower counts measured
by \citeN{DSP}. \citeANP{1998ApJ...505...50B} claim that the aperture
corrections employed by \citeN{DSP} resulted in an overestimation of
the depths of the latter survey by $\sim 0.5^m$; our results support
this interpretation.

It is worth noting that in our last bin, as in
\citeANP{1998ApJ...505...50B}, our counts are only approximately $50\%$
complete and therefore subject to large corrections. Furthermore, given
the small areas of surveys at this magnitude level (where areas covered
are typically $\sim 1$ arcmin$^2$) it is also possible that cosmic
variance could account for the large observed field-to-field variations
between the different groups.

Considering the slope of our number counts, we derive a value of
$d(\log N)/dm~=~0.37~\pm~0.03$ ($1\sigma$ errors) over the magnitude
range covered by both wide and deep catalogues (namely, $K~\sim
16-23$).  This value is in good agreement with $d(\log
N)/dm~=~0.36~\pm~0.02$ found by \citeANP{1998ApJ...505...50B}. We do
not find the count slope for the deeper survey to be significantly
different for the shallower one (however, as pointed out by \citeN{GCW}
the compiled $K-$ band number counts show a slope change at $K\sim17$).

In Figure~\ref{fig:countscols.counts} we also plot the predictions of
two non-evolving galaxy count models: one with deceleration parameter
$q_0=0.5$ (solid line) and one with $q_0=0.05$ (dotted line). Both
models give a good fit the observations from $K\sim14$ to $K\sim22$.
These non-evolving model were computed from the $k-$ corrections of
\citeN{BC} and the $z=0$ luminosity function parameters given in
Table~\ref{tab:pars}. We compute $M^*_K$ by combining the tabulated
$M^*_B$ with the rest-frame $(B-K)$ colours. In our models we adopt the
same normalisation as in our previous papers. As we and others have
noted before \cite{1991MNRAS.249..498M,sh89}, normalising galaxy counts
at brighter ($B\sim16^m$) magnitudes requires that substantial amounts
of $B-$ band evolution must take place at relatively low ($z<0.1$)
redshift.  Conversely, by normalising at $18^m<B<20^m$ we allow a
better fit at intermediate magnitudes and reduce the amount of
evolution required to fit the counts at fainter magnitudes, but
under-predict counts at brighter magnitudes. Support for this high
normalisation has recently come from the 2-micron all-sky survey (2MASS;
\shortciteNP{1992robt.proc..203K}) which finds a significant underdensity in
the galaxy counts at $10^m<K<13^m$ and is discussed further in a
forthcoming paper (Shanks et al 1999).


Could substantial galaxy luminosity evolution be occurring at these low
redshifts?  Bright, $K<20^m$ samples are dominated by early-type
galaxies Morphologically-segregated number counts from the Medium Deep
Survey \cite{1994ApJ...437...67G} and the Hubble Deep Field \cite{WBD}
indicate that counts for elliptical galaxies at intermediate magnitude
levels ($18^m<I_{816W}<22^m$) follow the predictions of a non-evolving
model
\cite{1998ApJ...496L..93D,1995ApJ...453...48D,1995MNRAS.275L..19G}.
This would argue against an evolutionary explanation for the steep
slope observed in the range $10^m<K<14^m$. We are then faced with
finding an alternate explanation for the number counts under-density;
one suggestion comes for the findings of \citeN{MFS} and
\citeN{1997A&A...317...43B} who both show how scale errors (i.e.,
non-linearities) in the photographic APM photometry of
\citeN{1990MNRAS.247P...1M} could result in anomalously low galaxy
counts at brighter magnitudes in $B-$ selected surveys; however, the
$K-$ band surveys are carried out with electronic detectors, and such
effects are likely to be less significant (although the
\citeN{1986MNRAS.223...11M} survey was based on objects selected from
$B-$ band photographic plates). Another suggestion is that our galaxy
may reside in a locally under-dense region of the Universe \cite{sh89};
however, to produce the observed discrepancy in the number counts would
require that this void would be $\sim 150 h^{-1}$~Mpc, and have an
under-density of $\sim 30 \%$; fluctuations on this scale are difficult
to understand in terms of current cosmological models and also measured
large-scale bulk flows.  Some evidence for a local void has come from
studies of peculiar velocities of type Ia supernovae
\cite{1998ApJ...503..483Z} although the size of the void they detect
($\sim70 h^{-1}$~Mpc) is not large enough to explain the low number
counts. For time being, resolution of this issue awaits a better
determination of the bright end of the $B-$ and $K-$ number counts by
much larger surveys.

\subsection{Optical-Infrared Colours}
\label{sec:Optic-Infr-Colo}

\begin{figure*}
\centering
\centerline{\epsfxsize = \sizex
\epsfbox{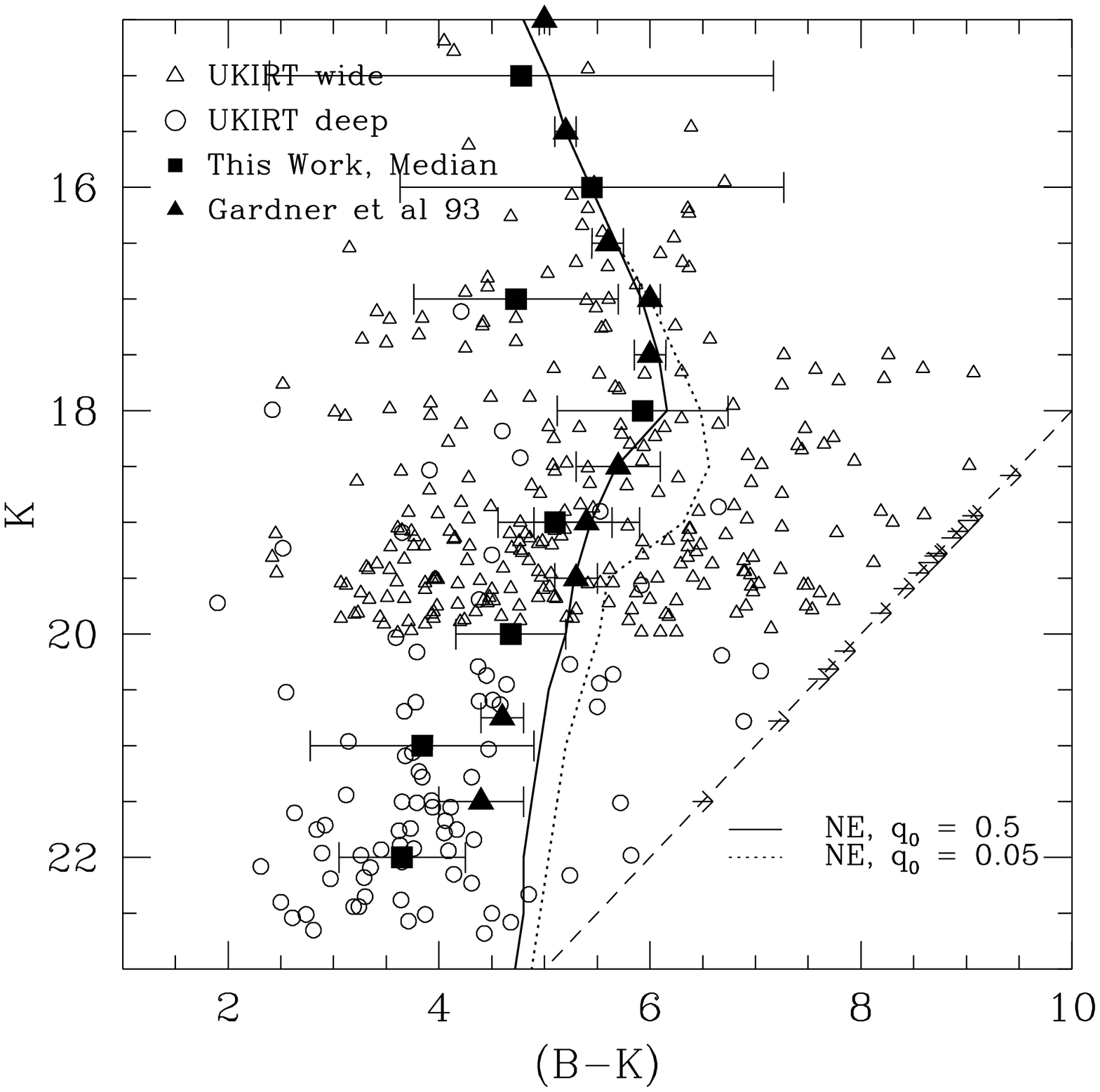}}
\caption{$K-$ magnitude against $(B-K)$ colour for galaxies in the
  wide survey (open triangles) and the deep survey (open circles). Also
  plotted are the median $(B-K)$ colours in one-magnitude bins for our
  survey (filled squares), with $1\sigma$ error bars calculated by a
  bootstrap resampling technique. Triangles show median colours from
  the compilation of Gardner et al. 1993. The dotted line represents
  the region of incompleteness; galaxies below this line have $B<28^m$.
  Drop-outs are plotted at the upper limit of their colours as
  right-pointing arrows. For comparison, predictions from two
  non-evolving models are plotted; one for $q_0 = 0.05$ (dotted line)
  and one for $q_0=0.5$ (solid line).}
\label{fig:medBK}
\end{figure*}

Figure~\ref{fig:medBK} shows the $(B-K)$ {\it vs} $K$ colour-magnitude
diagram for both samples for galaxies in both the ``wide'' (open
triangles) and the ``deep'' survey (open circles). Additionally the
median $(B-K)$ colour computed in one-magnitude bins is shown (filled
squares) and compared to the values given in \citeN{GCW} (filled
triangles). Error bars on the median colours were calculated using a
bootstrap resampling technique.

Incompleteness is represented by the dashed line; objects to the right
of this line have $B>28^m$. Several objects have been identified in
both surveys which are undetected in $B-$ but are detected in $K-$ and
these are plotted in the figure as right-pointing arrows and are
discussed in Section~\ref{sec:ExtRed}.  Considering the median $(B-K)$
colour we see that it increases steadily until around $K\sim 18^m$
where it reaches a maximum value of $\sim 6$.  After this, it sharply
becomes bluer and this blueward trend continues to the faintest limits
we have measured. Our median colours agree well with those in
\citeN{GCW}, even for brighter bins where the number of galaxies in our
sample are much smaller. We have also found our $K<18$ $(B-K)$
distributions to be in good agreement with the brighter survey of
\citeN{1997AJ....114..887S}. Finally, it is worth noting the extremely
large spread in colour in this diagram -- at $K\sim 20$ $(B-K)$ ranges
from $2$ to $9$.

Additionally, the predictions of the two non-evolving models shown in
Figure~\ref{fig:countscols.counts} are plotted. In general, the
$q_0=0.5$ model gives a slightly better fit to the observed colours, at
least until $K~\sim~20^m$; at fainter magnitudes, galaxies are
significantly bluer than both model predictions.

Another interesting feature in Figure~\ref{fig:medBK} is the apparent
deficit of galaxies with magnitudes in the range $K>21^m$
and colours from $(B-K) > 5$. The different areal coverages and
depths of the two surveys doubtless exacerbates this effect but its
presence is consistent with a population of galaxies which turns
rapidly blueward faintwards of $K~\sim~20$. We discuss the origin of
this feature in Section~\ref{sec:Medi-Colo-Counts}.

\begin{figure*}
\centering
\centerline{\epsfxsize = \sizex
\epsfbox{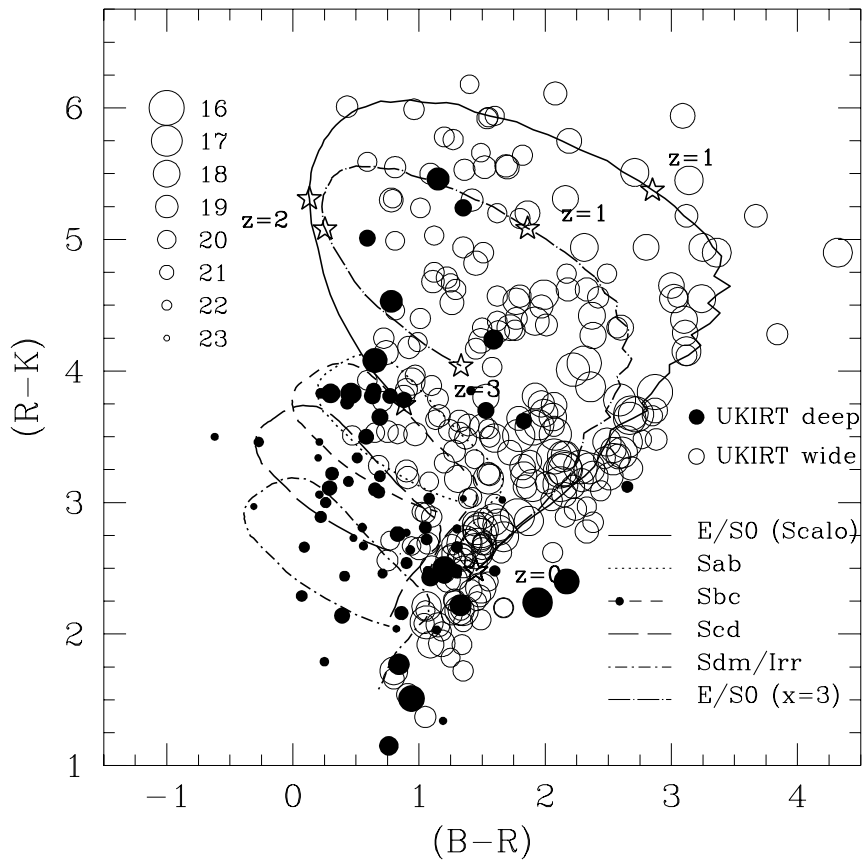}}
\caption{$(B-R)$ colour against $(R-K)$ colour for galaxies in the wide
  survey (open circles) and the deep survey (filled circles). Fainter
  galaxies are plotted with smaller symbols. Also shown are
  evolutionary tracks for the five classes of galaxy in our model, as
  computed from the population synthesis code of Bruzual \& Charlot
  (93). The solid line represents type E/S0, the dotted line Sab and
  the dashed, long dashed and dash-dotted lines types Sbc, Scd and Sdm
  respectively. The dot-dashed line shows the path of of an elliptical
  galaxy which has an IMF slope, x, of 3. For the elliptical tracks,
  the labelled stars indicate the redshift.  }
\label{fig:Col:Col}
\end{figure*}

In Figure~\ref{fig:Col:Col} we plot $(B-R)$ {\it vs.} $(R-K)$ colour
for all the galaxies in the wide survey (open circles) and the deep
survey (filled circles). Fainter galaxies are plotted with smaller
symbols. The ``deep'', $K<22.75^m$ sample is much bluer in both $(B-R)$
and $(R-K)$ than the ``wide'' than the brighter, $K<20^m$, sample,
which reaches $(R-K)\sim6$. We also plot colour-colour tracks
representing the different galaxy classes in our evolutionary models.
We defer discussion of these tracks until the following section.

\section{Interpretation and Modelling}
\label{sec:Interpr-Modell}

\subsection{Outline of the Models}
\label{sec:Outline-Models}

To interpret our results we investigate variants of a pure luminosity
evolution (PLE) model in which star-formation increases exponentially
with look-back time. Earlier versions of these models are discussed in
our previous papers \cite{1991MNRAS.249..498M,MSFR,MSCFG}. In this
paper we consider cosmologies in which $\Lambda=0$ and take
$H_0=50$~kms$^{-1}$Mpc$^{-1}$, although changing the value of $H_0$
does not affect any of the conclusions of this paper. Two values of the
deceleration parameter, $q_0$, $q_0 = 0.05$ and $q_0 = 0.5$, are
adopted, corresponding to open and flat cosmologies respectively.
(Zero-curvature cosmologies with $\Omega=0.3$ and $\lambda_0=0.7$ are
almost identical to our $q_0=0.05$ model because co-moving distance and
look-back time as a function of redshift are almost identical to that
found in the zero-$Lambda$ models, at least to $z\sim2$
\cite{1984ApJ...284..439P}). The input parameters to our models (given
in Table~\ref{tab:pars}) consist of observed {\it local\/} galaxy
parameters (namely, rest-frame colours and luminosity functions) for
each of the five morphological types (E/S0, Sab, Sbc, Scd and Sdm) we
consider in our models. These morphological types are divided into
elliptical (E/S0) and spiral (the remainder) and these two classes are
each given a separate star-formation history. We could, in principle,
sub-divide the spirals into different morphological types each with
different star formation histories but for simplicity we do not;
$(k+e)$ corrections for the different types are fairly similar to each
other in these models in any case . Instead, taking a Sbc model as
representative of all types we produce the other types by normalising
the Sbc track to the observed rest-frame $(B-K)$ colours from Table
\ref{tab:pars}.  As we have already explained in
Section~\ref{sec:countscols.Kband} our $\phi^{*}$'s are chosen to match
the galaxy counts at $B\sim18-20$ and we seek to explain the low number
counts at bright magnitudes from a combination of photometric errors
and anomalous galaxy clustering, rather than substantial evolution at
low redshift.  Our models also include the effects of the
Lyman-$\alpha$ forest, and, for spiral types, dust extinction
corresponding to the Large Magellanic Cloud as described in
\citeN{Pei}.

PLE models have difficulty in correctly reproducing the observed $K-$
band redshift distribution. Even essentially passive evolution
over-predicts the numbers of galaxies seen at $1<z<2$, and this fact
has been used by several authors (most recently
\citeN{1998MNRAS.297L..23K}) to argue for the hierarchical,
merger-driven galaxy formation models in order to reduce the mean
redshift of $K-$ selected galaxy redshift distributions. However, in
\citeN{MSCFG} we showed how PLE models {\it could\/} produce a $K-$
band redshift distribution compatible with the observations of
\citeN{CSH2} by assuming a very steep slope ($x=3$) for the stellar
initial mass function (IMF) in E/S0 types. This steep, low-mass-star
dominated IMF reduces the amount of passive evolution observed in $K-$
and allows us to reproduce the observed redshift distribution, which is
close to the predictions of the non-evolving model. It should be noted
however, that if the incompleteness in the \citeN{CSH2} n(z) were
entirely due to $z>1$ galaxies then the deficit relative to the PLE
model (with standard IMF) is only $\sim 3\sigma$ from $\sqrt N$
statistics alone and is likely to be less significant when galaxy
clustering is taken into account. Conversely, the $B-$ band redshift
distribution from the same survey does have a large fraction of high
redshift ($z>1$), high luminosity ($L>L^*$) galaxies. However, this is
not a problem for our models. Our steep luminosity function slope for
spiral types combined with a small amount of internal dust extinction
(corresponding to $A_B=0.3^m$ at $z=0$) allows us to reproduce this
result.

Although the resolution of the dataset discussed in this paper is too
low to extract useful size information for the faint galaxy population,
we mention in passing that observed faint galaxy sizes are compatible
with these models, models, as we have demonstrated in a recent paper
\cite{1998yugf.conf..102S}. We have used a modified version of Howard
Yee's PPP code to produce a simulated HDF frame which contains galaxies
with parameters based on the output of our PLE models. For this image,
spiral galaxies were generated assuming that luminosity evolution only
affects disk surface brightness and not size, combined with Freeman's
\citeyear{1970ApJ...160..811F}, law to relate disk size to absolute
magnitude at z=0. For bulges, we adopted the the diameter-magnitude
relation of \citeN{1990ApJ...361....1S}.  By applying our photometry
package to this image we account for selection effects in a realistic
manner. We find that the galaxies in the simulation have sizes not
dissimilar from those measured in the HDF, in part because many are
intrinsically small galaxies seen well down the LF.

\begin{figure}
  \begin{center}
    \centerline{\epsfxsize = 12cm
      \epsfbox{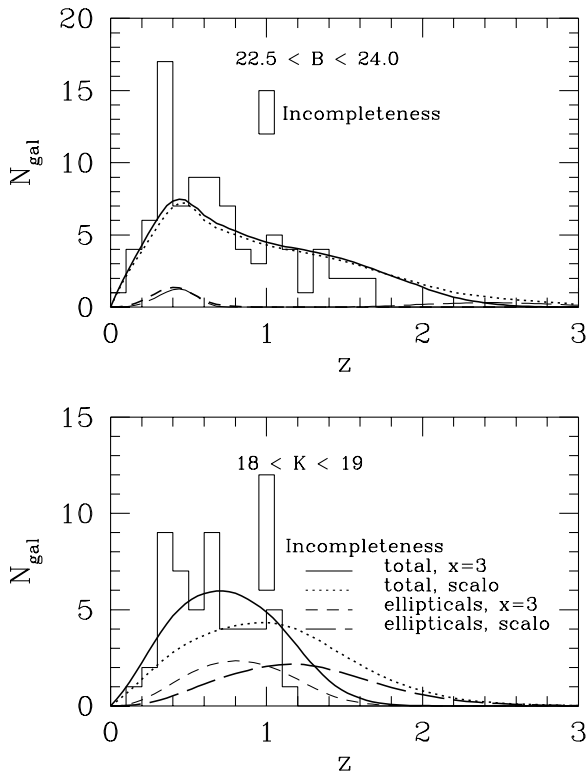}}
    \caption{Predicted redshift distributions for both ``Scalo'' ($\tau 
      = 1.0$~Gyr) and ``x=3'' ($\tau=2.5$~Gyr) models (dotted lines and
      solid lines) compared with the results of the $K-$ (lower panel)
      and $B-$ (upper panel) redshift distributions taken from Table 1
      of Cowie et al. 1996. Incompleteness is represented by the open
      boxes. Also plotted are the predicted redshift distributions for
      elliptical types (long dashed and short dashed lines) for the
      ``Scalo'' and ``x=3'' models respectively.}
    \label{fig:redshift.dndz}
  \end{center}
\end{figure}

\begin{figure}
  \begin{center}
    \centerline{\epsfxsize = 9cm
    \epsfbox{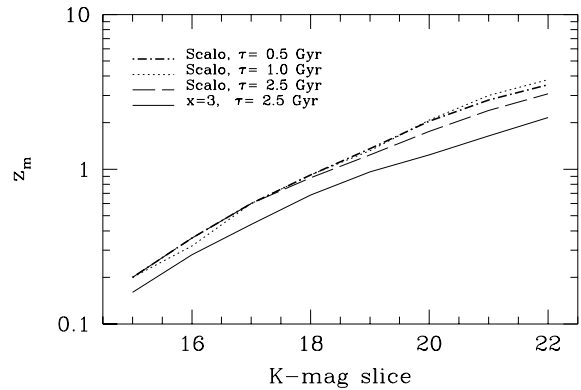}}
  \caption{The median redshift, $z_{med}$ of elliptical galaxies in our model
    populations for the magnitude slices used in
    Figure~\ref{fig:BKhists} and Figure~\ref{fig:IKhists}.These
    galaxies have star-formation histories.}
\label{fig:medzed}
  \end{center}
\end{figure}

In Figure~\ref{fig:redshift.dndz} the effect which changing the IMF
slope for early types has on predicted redshift distributions is
illustrated.  The dotted line shows a model computed using $\tau =
1.0$~Gyr and a Scalo \cite{1986FCPh...11....1S} IMF, whereas the solid
line shows the $x=3$ model which has $\tau=2.5$~Gyr. Also plotted are
the redshift distributions taken from Table 1 of \citeN{CSH2}. Objects
which were marked as unidentified in each band in this table are shown
in their respective incompleteness boxes. We have normalised both model
predictions to the total number of {\it observed\/} objects (i.e.,
unidentified objects are included).  We take $q_0 = 0.05$ and $z_f=6.4$
in all our models. The star-formation e-folding time $\tau$, is
$9.0$~Gyr for all the spiral types.  The $x=3$ model, as it dominated
by dwarf stars produces {\it much smaller\/} amounts of passive
evolution in the $K-$ band and as a consequence the median redshift of
galaxies in this model is much lower than in the Scalo model, which
produces an extended tail in the redshift distribution. Notice the
effect of changing the IMF slope in the $B-$ band is negligible.

This point is further illustrated in Figure~\ref{fig:medzed} where we
show the median redshift of elliptical galaxies with a range of
star-formation histories for several $K-$ selected magnitude slices.
This panel shows that, in all all magnitude ranges, elliptical galaxies
in the ``Scalo'' model have a higher median redshift than in the
``x=3'' model, irrespective of $\tau$ (demonstrating that the
differences in \ref{fig:redshift.dndz} are due to changing the IMF and
not $\tau$). This point is discussed further in
Section~\ref{sec:Medi-Colo-Counts}.

\begin{table*}
\begin{center}
\caption {Luminosity function parameters and rest-frame colours used in all models}
\begin{tabular}{cccccc}
Parameter / Type  & E/S0 & Sab & Sbc & Scd & Sdm\\\\
$\phi^{*}$ (Mpc$^{-3}\times 10^{-4}$) & $10.2$ & $5.09$ & $6.82$ & $3.0$ & $1.5$\\
$\alpha$ & $-0.7$ & $-0.7$ & $-1.1$ & $-1.1$ & $-1.5$\\
$M^*_B$ & $-21.0$ & $-21.0$ & $-21.32$ & $-21.44$ & $-21.45$\\
$(B-K)$ & $3.93$ & $3.78$ & $3.51$ & $2.90$ & $2.26$\\
$(I-K)$ & 1.57 & 1.63 & 1.7 & 1.47 & 1.0\\
\label{tab:pars}
\end{tabular}
\end{center}

\end{table*}

\begin{figure}
  \begin{center}
    \centerline{\epsfxsize = 12cm
    \epsfbox{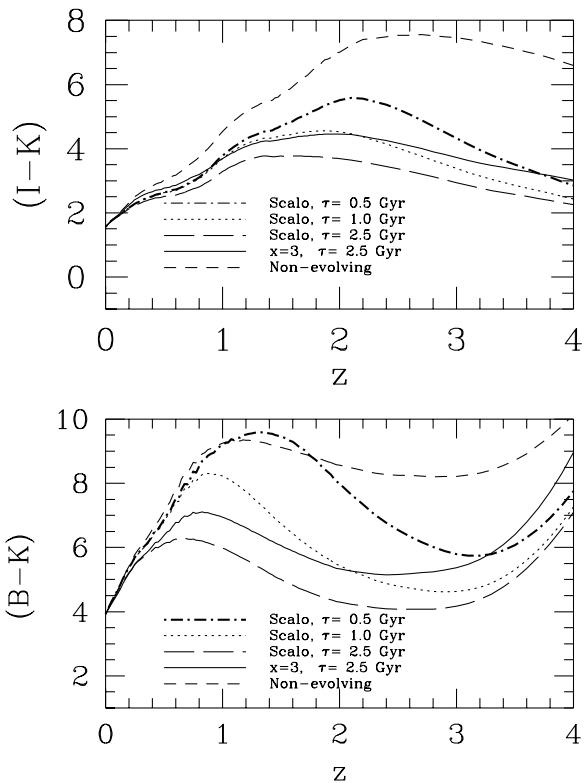}}
    \caption{{\bf a:} (lower panel) $(B-K)$ colour against redshift, $z$,
      for an elliptical galaxy with an $x=3$ IMF and $\tau=2.5$~Gyr
      (solid line), and for the Scalo IMF and with three values of
      $\tau$; $\tau=0.5$~Gyr (dot-dashed line), $\tau=1.0$~Gyr (dotted
      line) and $\tau=2.5$~Gyr (long dashed line). Also shown is a
      non-evolving model (short dashed line). {\bf b:}(upper panel) As
      for the lower panel, but with $(I-K)$ colour.  }
\label{fig:colzedmed}
  \end{center}
\end{figure}

\begin{figure}
  \begin{center}
    \centerline{\epsfxsize = 12cm
      \epsfysize = 12cm
    \epsfbox{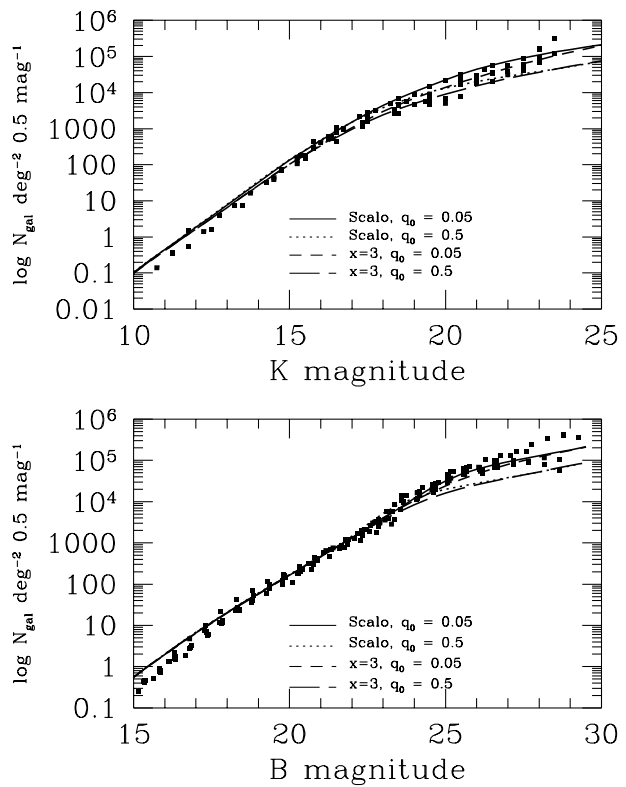}}
    \caption{Galaxy number counts for two evolutionary
      models -- the ``scalo'' model, (solid and dotted lines), and the
      ``x=3'' model, (short dashed and long-dashed), for low and high
      values of $q_0$ respectively. $K-$ band counts are shown in the
      upper panel; $B-$ band ones in the lower one.}
    \label{fig:slope.removed}
  \end{center}
\end{figure}

In Figure~\ref{fig:colzedmed} we illustrate the effect which varying
the IMF slope and $\tau$ has on the E/S0 colour-redshift relation. The
solid, dotted and long-dashed lines shows Scalo IMF tracks with $\tau =
0.5$~Gyr, $\tau=1.0$~Gyr and $\tau = 2.5$~Gyr respectively, whereas the
long dashed lines shows an $x=3$ model with $\tau = 2.5$~Gyr. As is
apparent from the plot, at $z\sim1$ $(B-K)$ colour depends {\it very
  sensitively\/} on the assumed value of $\tau$ and the slope of the
IMF\@. For a given IMF, longer $\tau$'s cause the peak in the
$(B-K)$-$z$ relation to shift to progressively higher redshifts, a
point we will return to in Section~\ref{sec:Medi-Colo-Counts}. Finally,
the $x=3$ IMF model track, as it is dominated by low-mass stars, is in
general redder (for a given value of $\tau$) than the Scalo IMF track.
We note that the single-burst models of \citeN{1994ApJS...95..107W}
have shown that the amount of passive evolution in the rest frame $I-$
and $R-$ bands is sensitive to metallicity variations; consequently at
$z=1-2$ the predicted $K-$ magnitude from our essentially single-burst model
is also expected to be approximately independent of metallicity.
Therefore reasonable changes in metallicity seem unlikely to change the
rate of passive evolution at $K-$ which, in the context of these models,
only appears to depend on the slope of the IMF."

We now return to Figure~\ref{fig:Col:Col} where we have over-plotted on
our $(B-R)-(R-K)$ diagram evolutionary tracks for each of the galaxy
types in our model. The dot-dashed and solid lines shows the path of an
elliptical, whereas the remaining tracks show the four spiral types in
the models. On the elliptical tracks, the labelled stars indicate the
redshift at selected intervals. Generally, there is good agreement
between the model tracks and the observed colours. The low redshift
($0.5 < z < 0.0$) elliptical track is populated by the brightest
galaxies in our sample. Generally, we do not find a large fraction of
objects have colours which cannot be reproduced by the tracks (in
contrast with the findings of \citeN{MDS97} who claimed there was a
significant population of objects whose colours could not be reproduced
by the models). There are {\it some\/} objects which appear to be
redder than the tracks, but these do not constitute a significant
fraction of the observed galaxy population. The broad range of $(B-K)$
colours of objects with $(R-K) > 4$ (which according to the models
should all be E/S0 galaxies) is a real effect which we interpret as
evidence of the diverse range of star-formation and metallicities
histories present in the elliptical population.

\subsection{Counts and Colour Distributions}
\label{sec:Medi-Colo-Counts}

In Figure~\ref{fig:slope.removed} we present a compilation of all
published galaxy number counts in $B-$ and $K-$ bandpasses, as well as
the predictions from our two evolutionary models. 

What is immediately apparent is that both the models discussed provide
a good fit to the number counts over a wide magnitude range for the low
$q_0$ case. Furthermore, in both $B-$ and $K-$ bandpasses at faint
magnitudes the differences between the Scalo and ``x=3'' models is {\it
  much smaller\/} than the differences between the high- and low-$q_0$
models. It is also apparent, as has been noted by many others
\cite{MSCFG,1997ApJ...488..606C,BR} that, unless another population
(such as bursting dwarfs) is added to the models then the $q_0=0.5$
model under-predicts the observed $B-$ band galaxy counts. In the $K-$
band this model is marginally excluded at the faintest magnitudes.

Number counts are a coarse test of any model; the extremely deep
optical photometry covering the Herschel Field is one of main virtues
of this work. So, to test our models more stringently we now
investigate our observed optical-infrared colour distributions. We
confine our discussion to the $q_0=0.05$ case as these models provide 
a better fit to the counts, although the colour distribution for the
$q_0=0.5$ models will be very similar.

\begin{figure*}
  \centering \centerline{\epsfxsize = \bigx
    \epsfbox{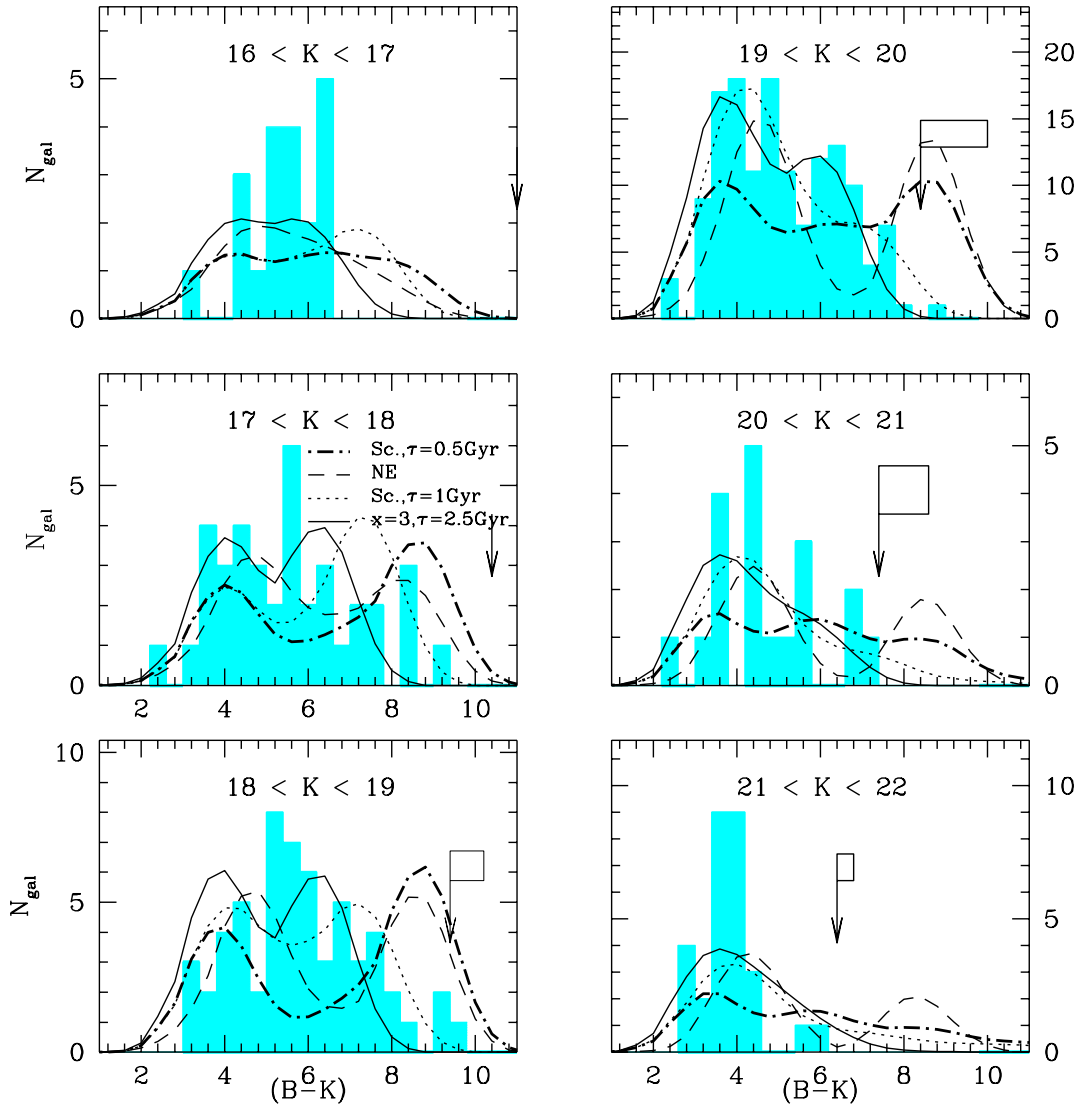}}
\caption{$K-$ selected $(B-K)$ colour distributions (solid histogram) 
  in six magnitude slices for both the wide (brightest four slices) and
  the deep survey (faintest two slices). Also shown are predictions
  from four models: `'x=3'' (solid line), ``Scalo'' (dotted and
  dot-dashed lines) and ``non-evolving'' (long dashed line). The two
  ``Scalo'' models have differing star-formation histories; the dotted
  and heavy-dashed line has $\tau=1.0$~Gyr whilst the heavy dot-dashed
  line has $\tau=0.5$~Gyr. The $x=3$ models has $\tau=2.5$~Gyr. The
  model predictions have been normalised to the {\it total\/} number of
  galaxies in each slice. In the fainter ($K>18^m$) bins this includes
  the $B-$ band non-detections discussed in Section~\ref{sec:ExtRed}
  and the number of such objects in each magnitude slice is represented
  by the incompleteness boxes. The position of the downward-pointing
  arrow represents the colour of the {\it reddest\/} object at the
  centre of each magnitude slice which could be detected; objects in
  the incompleteness boxes must lie rightwards of this arrow. For all
  models in this plot we adopt $q_0=0.05$.  }
\label{fig:BKhists}
\end{figure*}

In Figure~\ref{fig:BKhists} we compare in detail our model predictions
with our observations for $K-$ selected $(B-K)$ colour distributions
and compare these with the data (shown as the solid histograms) in six
one-magnitude slices from $K=16^m$ to $K=22^m$. The ``Scalo'' model is
shown as a dotted and a dot-dashed line and the ``x=3'' model is
represented by a solid line. All models have been convolved with a
conservative $0.3^m$ colour measurement error (overestimating errors in
the brighter magnitude slices). For the non-evolving model, as for the
evolving models, $k-$ corrections are calculated from the models of
\citeN{BC}.

\begin{figure*}
\centering
\centerline{\epsfxsize = \bigx
\epsfbox{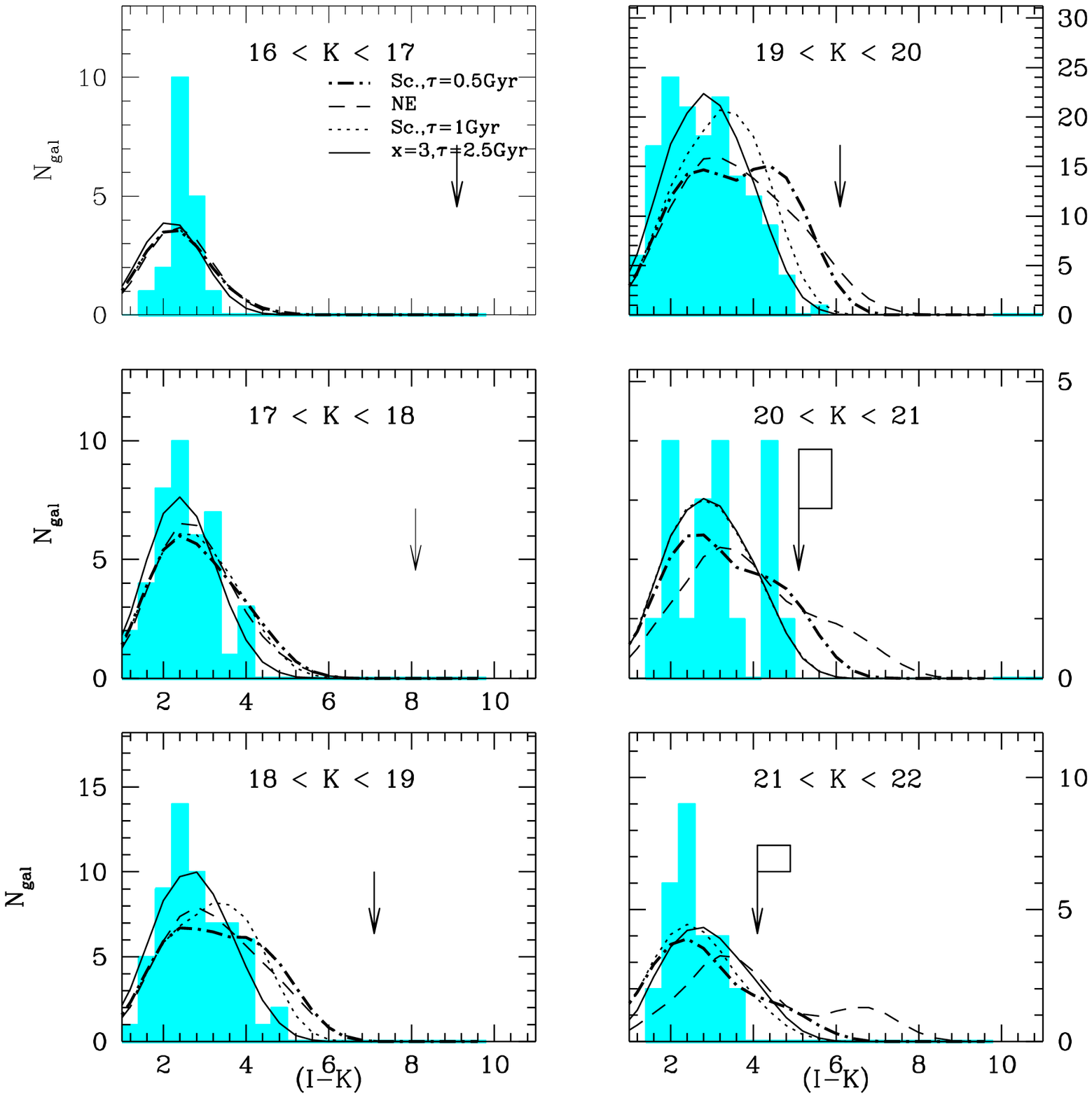}}
\caption{$K-$ selected $(I-K)$ colour distributions for 
  both the wide and deep surveys (solid histogram). All symbols are as
  in Figure~\ref{fig:BKhists}. }
\label{fig:IKhists}
\end{figure*}

Incompleteness in the histograms was determined by counting how many
galaxies had $K-$ magnitudes but were not detected in $B-$ in each
magnitude slice (and consequently have undetermined colours). The
colour of the {\it reddest\/} object which could be detected is
represented by the downwards-pointing arrow, which is plotted at the
colour corresponding to the $K-$ magnitude at the centre of each
magnitude slice (these objects have $B>28^m$).  The number of
non-detections is shown by the size of the box. Models have been
normalised to the total number of galaxies in each slice; in the latter
panels this includes $B-$ band non-detections.

The non-evolving model predictions (dashed line) rapidly diverge from
the data. Faintwards of the $K=17^m-18^m$ magnitude slice a bump
appears in the model colour distributions, corresponding to the
unevolved, red, $(B-K)\sim8$ elliptical population.  At approximately
the same magnitude limit and at bluer colours (this time at $(B-K) \sim
4$) a second peak becomes apparent.  This corresponds to the model
spiral population, and at fainter magnitudes this peak becomes the more
prominent of the two. This is a consequence of our adopted luminosity
function parameters and morphological mix. 

Qualitatively, the evolutionary models follow the pattern of the
non-evolving model, though the location and amplitude of the E/S0 peaks
depend sensitively on the choice of IMF and $\tau$.  On the blueward
side of the distributions, the spiral model colours appear to be a good
fit to the observations.  Referring to Figure \ref{fig:colzedmed} it is
apparent that beyond $z\sim0.3$, the $(B-K)$ colour for the
$\tau=0.5$~Gyr E/S0 model is redder than the $\tau=1.0$~Gyr model;
furthermore, as the median redshift of both populations is almost the
same (Figure~\ref{fig:medzed}), changes in the colour distributions are
almost completely a product of differences in the $(B-K)$-$z$ relation.
In general, \citeN{BC}-type models for elliptical galaxies with short
$\tau$'s predict a $(B-K)-z$ relation in which galaxies are at their
reddest at $z\sim 1$, and become bluer thereafter until $z\sim3$ at
which point they turn redder once more as a result of Lyman-$\alpha$
forest reddening.  The position of this turn-over depends on the
assumed $\tau$, the IMF and the redshift of formation $z_f$.

If we now we consider the ``x=3'', $\tau=2.5$~Gyr model, we
see that its colour distributions are in general {\it bluer\/} than
those predicted from the ``Scalo'' $\tau = 0.5$ or $\tau = 1.0$~Gyr
models.  This is partly a consequence of the longer $\tau$ adopted in this
model, but also of the fact that (as we have seen in
Figure~\ref{fig:medzed}) galaxies lie at a lower redshift than the
``Scalo'' model and consequently are sampling a bluer region of the
$(B-K)$-$z$ relation. So, despite the $x=3$ model being low-mass star
dominated, the larger value of $\tau$ allows us to produce colours
which are blue enough to match the observations.

Returning to Figure \ref{fig:BKhists} we can now understand the origin
of the differences between the $(B-K)$ colour distributions predicted
by the models.  Considering first the brighter magnitude slices in
Figure~\ref{fig:BKhists}, at $K=17.0^m-18.0^m$ we see that the
``Scalo'' $\tau=1.0$~Gyr model predicts a peak at $(B-K) \sim 7$ which
appears in the ``x=3'' model at $(B-K)\sim6$. At this magnitude,
elliptical galaxies in the ``Scalo'' model have $z_{med}\sim0.6$,
whereas for the ``x=3'' model $z_{med}\sim~0.4$ (from
Figure~\ref{fig:medzed}); as a consequence ``Scalo'' model colours are
redder than ``x=3'' ones.  Furthermore, the essentially
passively-evolving Scalo $\tau=0.5$~Gyr model contains insufficient
recent star-formation to shift the peak at $(B-K)\sim9$ blueward; as we
have seen from our considerations of the non-evolving model, this peak
is from unevolved ellipticals. At fainter ($K>20^m$) magnitudes, both
the $x=3$ model and the Scalo $\tau =1.0$~Gyr model turn sharply
blueward; the Scalo $\tau=0.5$~Gyr model colour distribution, however,
has a extended tail which reaches $(B-K)\sim 9-10$.

It is clear from Figure \ref{fig:BKhists} that either a Scalo IMF with
a $\tau = 1.0$~Gyr or an $x=3$ IMF with $\tau=2.5$~Gyr model
reproduces the main features of the data histograms, including the
broadening of the colour distribution at the $K<18^m$ slice, and its
subsequent narrowing and blueward trend in the fainter slices.
Furthermore, as we will see in the following Section, these models are
consistent with the redwards limits of our data.  Of these two models,
the $x=3$, $\tau=2.5$~Gyr model as commented in
Section~\ref{sec:Outline-Models}, gives a much better fit to the
observed $K-$ band redshift distributions.  As is apparent from
Figures~\ref{fig:medzed} and~\ref{fig:colzedmed} distinguishing between
the effects of changes in IMF slope or the star-formation history on
the colour-redshift relation is difficult. Making this choice requires
additional information such as the redshift distribution of $K-$
selected galaxies.

Finally, these models also allow us to understand the deficit of
galaxies in the colour-magnitude relation at $(B-K)>5$ and $K>21^m$
which we commented upon in Section~\ref{sec:Optic-Infr-Colo}.  Firstly,
as a consequence of our choice of luminosity function parameters
elliptical counts turn over at $K\sim20$ and spiral types dominate
faintwards of this.  Secondly, from Figure \ref{fig:medzed} we see that
at $K\sim21$ E/S0's are at $z\sim1$; from these redshifts to $z>3$
Figure \ref{fig:colzedmed} indicates that elliptical colours turn
sharply blueward. In combination, these factors produce the rapid
blueward trend observed at $K>21^m$ and the absence of redder objects,
which happen in all our models regardless of our choice of IMF or
$\tau$.

\subsection{$(I-K)$ distributions and extremely red galaxies}
\label{sec:ExtRed}
In the previous section we discussed the broad characteristics of the
$K-$ selected $(B-K)$ colour distribution and showed how these features
can be understood in terms of our models; in this section we turn our
attention to very red objects which are at the outskirts of the colour
distributions. The red colours of these objects imply they are E/S0
galaxies at moderate ($z>1$) redshift and consequently it is
interesting to see if their observed number density is consistent with
current galaxy formation scenarios. As previous authors
\cite{AB,MDS97,1997Natur.390..377Z,1994ApJ...434..114C} who focussed
their attentions upon these galaxies discussed them in terms of $K-$
selected $(I-K)$ distributions, here we introduce our observations in
these bandpasses. Following these workers we define an ``extremely red
galaxy'' as an object with $(I-K)>4$; from Figure~\ref{fig:colzedmed}
we see that at $(I-K)\sim4$, $z\sim1$. In this section we will make
comparisons with models generated using the same set of parameters as
the $K-$ selected $(B-K)$ distributions discussed previously. In
interpreting $(I-K)$ colours one faces some significant differences to
$(B-K)$ observations; as is apparent from Figure \ref{fig:colzedmed},
$(I-K)$ colour is relatively insensitive to the amount of
star-formation; at $z\sim1$ the difference in $(I-K)$ between
elliptical galaxies with a Scalo IMF and $\tau=0.5,1,2.5$~Gyr is $0.5$;
by comparison, at this redshift $(B-K)$ ranges from $\sim3$ to $\sim5$.
This difference is reflected in the tightness of the $K-$ selected
$(I-K)$ distributions plotted in Figure \ref{fig:IKhists}. This
insensitivity makes $(I-K)$ distributions relatively poor probes of the
elliptical star-formation history.  However, it is precisely this
behaviour which makes $(I-K)$ colours useful in detecting
high-redshift, evolved ellipticals; Figure~\ref{fig:colzedmed} shows
that until $z\sim2$ $(I-K)$ colour is approximately proportional to
$z$, and is relatively insensitive to variations in $\tau$ or the IMF.

Turning to work from the literature, \citeN{1994ApJ...434..114C}, in a
$5.9$~arcmin$^2$ survey, detected $13$ objects with $(I-K)>4$ to
$K<20.9^m$, and concluded that these galaxies do not dominate the faint
$K>20^m$ population. They also concluded that less that $10\%$ of
present day ellipticals could have formed in single-burst events. In
contrast, \citeN{MDS97} found 8 galaxies with $(I-K)>4$ and $K<22^m$
over a small, $2$~arcmin$^2$ area. They also isolated a population of
objects with blue optical ($(V-I)<2.5$) and red near-infrared colours
and argued that these colours could not be reproduced with
\citeN{BC}-type models.  Although we do not have $V-$ band photometry
the top middle and top left of our Figure~\ref{fig:Col:Col} roughly
corresponds to the regions highlighted in the colour-colour plot shown
in Figure 9 of \citeANP{MDS97}. These ``red outlier'' objects, as
\citeANP{MDS97} describe them, viewed in the context of the models
discussed here are most likely elliptical galaxies at $1<z<2$. Finally,
both \citeN{1997Natur.390..377Z} and \citeN{AB} investigated extremely
red objects in the Hubble Deep Field (HDF).
\citeN{1997Natur.390..377Z}, using publicly-available Kitt Peak
near-infrared imaging data covering the $5$~arcmin$^2$ of the HDF found
that to a $50\%$ completeness limit of $K<22^m$ there were $\sim 2$
objects with $(V_{606}-K) >7$.  \citeN{AB}, in a much larger $\sim
60$~arcmin$^2$ infrared survey comprising both the HDF and the HDF
flanking fields found 12 galaxies with $(I-K)>4$ and $19<K<20$.

\begin{table*}
\caption{ Numbers of objects with $(I-K)>4~$per~arcmin$^{2}$ with $\pm
  1\sigma$ errors. The survey area in arcmin$^{2}$ is shown in
  parentheses. In the deep survey ($20<K<22$) there are 7 objects
  which are undetected in $I$ and therefore only lower limits can be
  placed on their colours. They are included in their respective bins.
  The number quoted from  Moustakas et al.~(1997) covers the magnitude 
  range $20<K<22$.}
\begin{center}
\begin{tabular}{lrlrlrl}
Author&\multicolumn{2}{c}{$19<K<20$}&\multicolumn{2}{c}{$20<K<21$}&
\multicolumn{2}{c}{$21<K<22$}\vspace{1mm}\\
Model ($x=3$)&\multicolumn{2}{c}{$0.3$}&\multicolumn{2}{c}{$1.4$}&
\multicolumn{2}{c}{$1.8$}\vspace{1mm}\\
Model (scalo, $\tau=1.0$~Gyr)&\multicolumn{2}{c}{$0.6$}&
\multicolumn{2}{c}{$1.4$}&\multicolumn{2}{c}{$1.2$}\vspace{1mm}\\
Model (scalo, $\tau=0.5$~Gyr)&\multicolumn{2}{c}{$1.1$}&
\multicolumn{2}{c}{$3.1$}&\multicolumn{2}{c}{$2.8$}\vspace{1mm}\\
This work&$0.5^{+0.1}_{-0.1}$&(47)&$3.9^{+2.1}_{-1.4}$&(1.8)&
$1.1^{+1.5}_{-0.7}$&(1.8)\vspace{1mm}\\
\citeN{AB}&$0.2^{+0.1}_{-0.1}$&(62)&$1.2^{+0.5}_{-0.4}$&(7.8)&
\multicolumn{2}{c}{$-$}\vspace{1mm}\\
\citeN{1994ApJ...434..114C}&$0.3^{+0.4}_{-0.3}$&(7.1)&$1.5^{+0.7}_{-0.5}$&(5.9)&
$0.3^{+0.5}_{-0.2}$&(5.9)\vspace{1mm}\\
\citeN{MDS97}&\multicolumn{2}{c}{$-$}&&\multicolumn{1}{r}
{$4.0^{+2.0}_{-1.4}$}&\multicolumn{1}{l}{(2)}&\\
\end{tabular}
\label{tab:ext-col-obj}
\end{center}
\end{table*}  

Table \ref{tab:ext-col-obj} presents a compilation of observations from
these papers as well as our current work. We also show the predictions
of the $x=3$ and Scalo $\tau=0.5$ model (our model which most closely
approximates a single burst), normalised to the total numbers of
objects each magnitude slice.  Generally, in the brighter bin
$(19<K<20)$ the only substantial comparison that we can make is with
\citeANP{AB}. We see a factor $\sim 3$ more objects than they do,
although this represents only a $2\sigma$ discrepancy. In the fainter
bin $20<K<22$ our numbers of red objects agree well with
\citeANP{MDS97}. Although our numbers are higher than
\citeN{1994ApJ...434..114C} in this range there is still no significant
discrepancy due principally to our small area.

\citeANP{AB} compared their observations to the predictions of a
passively-evolving $\tau=0.1$~Gyr model elliptical population, and
concluded that the numbers of $(I-K)>4$ objects observed disagreed with
the predictions of this model.  Our $\tau=0.5$~Gyr distribution (solid
dot-dashed line) in Figure~\ref{fig:IKhists} is quite similar to this
model, and in the $19<K<20$ magnitude slice we can see from
Table~\ref{tab:ext-col-obj} that it does indeed produce a large number
of objects with $(I-K)>4$ which are not seen in our observations.  We
therefore conclude, as \citeANP{AB} did, that the low numbers of
objects with $(I-K)>4$ observed at $19<K<20$ disfavour non-evolving and
passively evolving models (the constraints on the models which can be
obtained from the distributions at fainter magnitudes are less
significant). However, we note that the $x=3$, $\tau=2.5$~Gyr and
$\tau=1.0$~Gyr Scalo models do correctly reproduce the observed colour
distributions. This is because both these models contain enough
on-going star-formation to move galaxy colours sufficiently blueward to
match the observations.

Lastly, could a merging model, like those adopted by
\citeN{1998MNRAS.297L..23K}, produce these results?  It is not possible
to rule this model out from the redshift data and it may even be said
that the difficulty that models with standard Scalo IMF's have in
fitting the $K<19$ redshift distribution is evidence in favour of the
merging scenario. We therefore consider whether the $K-$ selected
$(B-K)$ and $(I-K)$ colour distributions presented here can
discriminate between a PLE $(x=3)$ model and a merging model. The $K-$
selected $(B-K)$ distributions in Figure~\ref{fig:BKhists} can be
understood in terms of either scenario; the red galaxies that are
missing at $17<K<22$ may either be due to them becoming blue at high
redshift due to continuing star-formation or passive evolution -- or
because they are fainter than expected due to de-merging (although our
lack of red galaxies even at the faintest limits implies either several
magnitudes of fading or that the pre-merger components are blue).
However, as we have already noted the $(I-K)$-$z$ relation for E/S0
galaxies is insensitive to changes in $\tau$ or the IMF.  Furthermore,
in the range $18<K<20$ some galaxies with the $I-K>4$ colour expected
of $z=1$ elliptical galaxies are detected (Figure \ref{fig:IKhists}).
Indeed the numbers of these galaxies we see is within a factor of $\sim
2$ of what is predicted on the basis of the Scalo, $\tau=0.5$~Gyr model
and in even better agreement with the $x=3$, $\tau=2.5$~Gyr models or
even a Scalo $\tau=1.0$~Gyr model. If the merging model is the
explanation of the large deficiency of red galaxies in $(B-K)$ compared
to passively evolving models, then it might be expected that a similar
deficiency should be seen in $(I-K)$.  Since the deficiency in $(I-K)$
is less, then this might be taken to be an argument {\it against}
merging and {\it for} the $x=3$ PLE model.

\section{Conclusions and Summary}
\label{sec:IntDisc}

In this paper we have presented the results of two near-infrared surveys to
$K\sim20$ and $K\sim23$ which cover our ultra-deep ($B\sim28$) optical
fields. We draw the following conclusions from this work:

\begin{description}
  
\item Our K number counts are consistent with the predictions of
  non-evolving models with $0\leq q_0 \leq 0.5$.
  
\item As previously noted by \citeN{CSH2},\citeN{MSCFG}, and
  \citeN{1998MNRAS.297L..23K} the $18<K<19$ $n(z)$ of \citeN{CSH2} is
  also well fitted by non-evolving models. However, passively evolving
  models with a Salpeter/Scalo IMF predict too many galaxies with
  $z>1$.  Dynamical merging is one possible solution to reduce the
  numbers of these galaxies but we have shown that a dwarf-dominated
  IMF for early-types could offer an alternative explanation.
  
\item Our $K-$ selected $(B-K)$ colour distributions display a strong
  bluewards trend for galaxies fainter than $K\sim20$, confirming
  results previously observed in the shallower surveys of \citeN{GCW}.
  
\item At brighter magnitudes ($K<20$~mag) our $K-$ selected $(B-K)$
  distributions indicate a deficiency of red, early-type galaxies at
  $z\sim1$ compared to the predictions of passively evolving models.
  This implies either a PLE model where star-formation continues at a
  low level after an initial burst or dynamical merging.
  
\item At fainter magnitudes ($20<K<22$) the continuing bluewards
  trend observed in $(B-K)$ can be explained purely in terms of
  passively evolving PLE models with no need to invoke any additional
  mechanisms.
  
\item Our observed numbers of $(I-K)>4$ galaxies at $K\sim20$ are lower
  than the predictions of passively evolving models or PLE models with
  a low level of continuing star-formation, suggesting that at least
  part of the larger deficiency observed in $(B-K)$ at $K\sim20$ may be
  due to star-formation rather than dynamical merging.
  
\item In the range $19<K<20$, where our statistical uncertainties are
  lowest, we detect $0.5\pm 0.1$ red galaxies arcmin$^{-2}$ with
  $(I-K)>4$.  We see a factor $\sim 3$ more objects than \citeANP{AB}
  do although this represents only a $2\sigma$ discrepancy.  The PLE
  models discussed here suggest these galaxies will have redshifts
  $1<z<2$. The numbers of these galaxies are consistent with
  the predictions of PLE models with small amounts of ongoing
  star-formation.
\end{description}

\section{Future Work}
\label{sec:Future-Work}
We have recently completed a $H-$ band survey of the Herschel field
using the $1024\times1024$ Rockwell HgCdTe array at the Calar Alto 3.5M
telescope. This survey is complete to $H<22.6^m$ and covers the entire
area of the Herschel Deep Field. Analysis of this dataset will be
presented in future papers.

\section{Acknowledgements}
\label{sec:Acknowledgements}
H.J.McCracken wishes to acknowledge financial support from PPARC and
the hospitality and generosity of Dr. Mariano Moles at IMAFF, Madrid
where an earlier version of this paper was written. Support for this
trip was also provided the University of Durham Rolling Grant for
Extragalactic Astronomy. This work was supported in part by a British
Council grant within the British/Spanish Joint Research Programme
(Acciones Integradas). NM acknowledges partial PPARC
support.

\end{document}